\documentclass[12pt]{article}
\usepackage{amsmath}
\usepackage{graphicx}
 \newcommand{\bea}{\begin{equation}}
 \newcommand{\eea}{\end{equation}}
 \newcommand{\ber}{\begin{eqnarray}}
 \newcommand{\eer}{\end{eqnarray}}
\begin{document}

\title{FOKKER-PLANCK EQUATION: LONG-TIME DYNAMICS IN
APPROACH TO EQUILIBRIUM}
\author{Himadri S. Samanta\footnote {e-mail:tphss@mahendra.iacs.res.in},
J. K. Bhattacharjee\footnote{e-mail:tpjkb@mahendra.iacs.res.in}\\
Department of Theoritical Physics\\
Indian Association for the Cultivation of Science\\
Jadavpur, Calcutta 700 032, India}

\maketitle
\begin{abstract}
We discuss the approach to equilibrium of systems governed by the
Fokker-Planck equation. In particular, we focus on
problems involving barrier penetration and the associated
Kramers' time. We also describe the connection between stochastic processes
and quantum mechanics. 
\end{abstract}
{\it{Keywards}}: Fokker-Planck equation; Approach to equilibrium; 
Barrier penetration

\section{Introduction}
In this article, we will discuss the approach to equilibrium
of a system with fluctuations. 
The simplest example of this is an overdamped particle 
with coordinate $x(t)$ at time $t$. This particle experiences 
an external potential $U(x)$.
The time evolution
of the position variable obeys the Langevin equation, which is  a stochastic 
differential equation:
\bea\label{1}
\frac{dx}{dt}=-\Gamma \frac{\partial U(x)}
{\partial x}+f(t)
\eea
where $\Gamma$ is the inverse friction constant and
$f(t)$ is the random or fluctuating force.
The distribution of $f(t)$ is Gaussion, and its correlation
function obeys

\bea\label{2}
<f(t_{1})f(t_{2})>=2\epsilon \delta(t_{1}-t_{2})
\eea 
where $\epsilon$ measures the amplitude of the noise term.

The equilibrium distribution for $x$ will be
proportional to $\exp[-U(x)/\epsilon]$ if the fluctuation-dissipation 
theorem holds, fixing $\Gamma =\epsilon$. It is
well known that if $P(x,t)$ is the time dependent probability
distribution for the random process described as $x(t)$, then
$P(x,t)$ satisfies the Fokker-Planck equation (Risken, 1984):

\bea\label{3}
\frac{\partial P(x,t)}{\partial t}=
\frac{\partial}{\partial x}
\Bigg (P\frac{\partial U}{\partial x}\Bigg )
+\epsilon \frac{\partial^{2}P}{\partial x^{2}}
\eea

It is clear that if $P=P_{0}$ $\propto \exp(-U/\epsilon)$,
then $\frac{\partial P}{\partial t}=0$ and this corresponds to the equilibrium
distribution. The approach to equilibrium from a non equilibrium state
will be the subject
of the ensuing sections.

\section{Kramers' Time}
In this section, we recall the approach to equilibrium in a system, 
whose time-dependent probability distribution is governed by a Fokker-Planck 
equation. In particular, we discuss the situation where the 
potential driving the dynamics is bistable. For the most part, we will 
talk about a one-dimensional potential.

A convenient way 
of handling the problem is to make the transformation 

\bea\label{4}
P(x,t)=\exp(-U/2\epsilon)\phi (x,t)
\eea 

For the new variable $\phi (x,t)$, one obtains 
the Schr\"{o}dinger-like equation 

\bea\label{5}
-\frac{\partial \phi}{\partial t} = -\epsilon \frac{\partial^{2}\phi}
{\partial x^{2}} + V(x)\phi
\eea
where

\bea\label{6}
V(x)=\frac{1}{4} \frac{U^{\prime}(x)^{2}}{\epsilon} - \frac{1}{2}
U^{\prime \prime}(x)
\eea

If $\phi_{n}(x)$ are a complete set of eigenfunctions of the Hamiltonian 
$H=-\epsilon \frac{\partial ^{2}}{\partial x^{2}}+V(x)$, then we can 
expand as follows:

\bea\label{7}
\phi (x,t)=\sum_{n} a_{n} \exp(-\lambda_{n}t) \phi_{n}(x)
\eea
where the $\lambda _{n}$ are the eigenvalues of $H$ given by

\bea\label{8}
H \phi_{n}=\lambda_{n}\phi_{n}
\eea
Further, the constants $a_{n}$ are determined dy 
$\phi(x,0)$, which is obtained from $P(x,t)$. 
It is clear from the form of H, which can be written as 
\bea\label{9}
H=\epsilon A^\dag A
\eea
with
\bea\label{10}
A=\frac{\partial}{\partial x}+\frac{U^{\prime}}{2\epsilon}
\eea
that the spectrum of $H$ is non negative and the lowest eigenvalue 
is zero with the corresponding eigenfunction given by
 $\phi_{0}=N \it{\exp}(-U/2\epsilon)$, where N is a 
normalization constant. It is clear from Eq.(\ref{7}) that 
as $t\rightarrow\infty $, only the term $n=0$ will servive 
$(\lambda_{0}=0)$ and the limiting value of $\phi(x,t)$ is 
$\exp(-U/2\epsilon)$, corresponding to equilibrium distribution 
$P_{eq} (x)\propto \exp(-U/\epsilon)$, as is obvious from Eq.(\ref{3}). 
Thus, in the Fokker-Planck equation, the approach to equilibrium 
is guaranteed. 

The approach to equilibrium occures on a fast time-scale if 
the potential has a single minimum, e.g. $U(x)=x^{2}$. 
In this case, $V(x)=\frac{x^{2}}{\epsilon}-1$, $\lambda _{n} 
= 2 n $ $(n=0,1,2,......)$ and Eq.(\ref{7}) reads as

\bea\label{11}
\phi(x,t)=a_{0}\phi_{0}(x)+a_{1}\phi_{1}(x) \exp(-2t)+
................
\eea 

The exponential term dies out on time-scales of $\it{O}(1)$ and 
the system, if started out from a non-equilibrium state, will evolve 
to the equilibrium state on a fast time-scale. Specifically, if we 
take the initial probability distribution to be $P_{0}(x)=\frac{1}
{(\pi \epsilon)^{1/2}} \exp(-(x-a)^{2}/\epsilon  )$, then 
$\phi(x,t=0)=1/(\pi \epsilon)^{1/2} \exp(-x^{2}/2\epsilon +ax/
\epsilon -a^{2}/2\epsilon)$, with the eigenfunction $\phi_{n}(x)$ 
given by $\phi_{n}(x)= ( \frac {1}{\epsilon^{1/2}
\pi^{1/2}2^{n}n!} )^{1/2} \exp(-x^{2}/2\epsilon) H_{n}
(x/\epsilon^{1/2})$. We find the expansion coefficients $a_{n}$ of 
Eq.(\ref{7}) to be given by $a_{n}=(a/\epsilon^{1/2})^{n}
\frac{1}{2^{n}n!}$ and the sum in Eq.(\ref{7}) is easily performed 
keeping in mind the generating function of Hermite polynomials to yield 

\bea\label{12}
P(x,t)=\frac{1}{(\pi \epsilon)^{1/2}} \exp \Bigg (-(x-ae^{-2t})^{2}/
\epsilon \Bigg )
\eea

\begin{figure}[t]
\begin{center}
\includegraphics[scale=0.5]{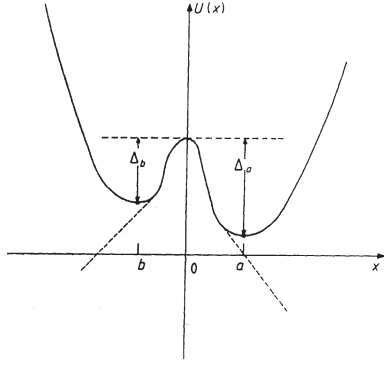}
\caption{\small{The general bistable potential in one dimension.}}
\end{center}
\end{figure}

We now turn our attention to the bistable potential shown in Fig.(1). 
At $t=0$, we choose a probability distribution centered sharply around 
$x=0$, which is clearly a non-equilibrium function. Because
 of the inverted oscillator potential near the center, the 
distribution will broaden out in the initial stages. Thereafter, 
in an intermidiate time zone called the Suzuki regime, the broadened 
central peak begins to split into two side peaks corresponding 
to the minima of $U(x)$ at $x=a$ and $x=b$. Finally the system 
enters the Kramers'(1940) regime where it is close to equilibrium but 
makes occasional large fluctuation due to the noise. This 
Kramers regime will be governed by the lowest eigenvalue 
$\lambda _{1}$ of the system. The inverse of $\lambda_{1}$ 
sets the time scale for attaining equilibrium. The eigenvalue 
$\lambda_{1}$ is exponentially close to the ground state value 
$\lambda_{0}=0$ and differs from it due to the tunneling in the 
three-well problem. 
A WKB calculation  of the eigenvalue was carried out by Caroli $\it{et} 
$ $\it{al.}$ (1979). 

Here we will show a variational calculation (Bhattacharjee and Banerjee, 1989) 
motivated by the 
pioneering work of Bernstein and Brown (1984). The observation 
of Bernstein and Brown was that the Hamiltonian 
$H=\epsilon A^{\dag}A$ and $\overline {H}=\epsilon AA^{\dag}$ 
are surpersymmetric partners having the same spectrum except 
for the ground state. While the ground state 
energy of $H$ is zero, the same can not be true for 
$\overline{H}$ because the wave function satisfying $A^{\dag}\phi =0$
cannot be normalized. The ground state of 
$\epsilon AA^{\dag}$ is then the first excited 
state of $\epsilon A^{\dag}A$. This is easily seen if we 
construct the matrix

\bea\label{13}
H_{s}=\left(\begin{array}{c}\begin{array}{cc}\epsilon A^{\dag}
A&0\\0&\epsilon AA^{\dag}\end{array}\end{array}\right)
\eea

The operator $Q$ with the representation $\left(\begin{array}{c}
\begin{array}{cc}0&0\\A&0\end{array}\end{array}\right)$ commutes with 
$H_{s}$ and hence $H_{s}$ has a degenerate spectrum. If $\psi$ 
is an eigenfuntion of $H_{s}$ with eigenvalue $\lambda$, then 
so is $Q\psi$. By inspection, one set of eigenvalues of $H_{s}$ 
can be written as $\psi_{n}=\left(\begin{array}{c}\phi_{n}\\0
\end{array}\right)$ since (see Eqs.(\ref{8}) and (\ref{9}))

\bea\label{14}
H_{s}\left(\begin{array}{c}\phi_{n}\\0\end{array}\right)=\epsilon 
\left(\begin{array}{c} A^{\dag}A\phi_{n}\\0\end{array}\right)
=\lambda_{n}\left(\begin{array}{c}\phi_{n}\\0\end{array}\right)
\eea 
Consequently, we must have 

\bea\label{15}
H_{s}Q\left(\begin{array}{c}\phi_{n}\\0\end{array}\right)=H_{s}
\left(\begin{array}{c}0\\A\phi_{n}\end{array}\right)=\lambda_{n}
\left(\begin{array}{c}0\\A\phi_{n}\end{array}\right)
\eea

Since $\left(\begin{array}{c}0\\A\phi_{n}\end{array}\right)=
\left(\begin{array}{c}0\\\epsilon AA^{\dag}A\phi_{n}\end{array}\right)$, 
it follows that

\bea\label{16}
\epsilon AA^{\dag}A\phi_{n}=\lambda_{n}A\phi_{n} 
\eea

Thus, $\lambda_{n}$ are the eigenvalues of $\epsilon AA^{\dag}$ with 
eigenfunctions $A\phi_{n}$. This will be true for all $\phi_{n}$ 
except $\phi_{0}$. The eigenvalue zero is nondegenerate. For all other 
$\phi_{n}$ we have $\left(\begin{array}{c}0\\A\phi_{n}\end{array}\right)$ 
and $\left(\begin{array}{c}\phi_{n}\\0\end{array}\right)$ 
as the degenerate eigenfunctions of $H_{s}$ and write $\phi_{n}$ 
are the eigenfunctions of $\epsilon A^{\dag}A$; the 
$A\phi_{n}$ are the eigenfunctions of $\epsilon AA^{\dag}$. 
To find the first excited state of $\epsilon A^{\dag}A$ 
(the one corresponding eigenfunction $\phi_{1}$), we need 
to find the ground state of $\epsilon AA^{\dag}$. Now that 
the problem is one of finding a ground state energy, this can be 
easily done by variational calculations.

We begin with the observation that the nonnormalizable wavefunction 
$\exp\Big(U(x)/2\epsilon \Big)$ satisfies 
$\epsilon AA^{\dag}\exp\Big (U(x)/2\epsilon \Big ) =0$. 
This motivates the choice of the wavefunction in the form 
$\exp\Big (\psi(x)/2\epsilon \Big )$ with

\bea\label{17}
\psi (x) = \begin{cases}
U(x), 
\qquad\qquad\qquad \text{for}\,\, 
b(1-\beta \epsilon^{1/2})\leq x\leq a(1-\alpha \epsilon^{1/2})
\\ 
U(a-a\alpha \epsilon^{1/2})-U^{\prime}
(a-a\alpha \epsilon^{1/2})(x-a+a\alpha \epsilon^{1/2}),  
\\\,\,\,\qquad\qquad\qquad\qquad\text{for}\, x\geq a(1-\alpha \epsilon^{1/2})
\\
U(a-a\beta \epsilon^{1/2})+U^{\prime}
(a-a\beta \epsilon^{1/2})(x-b+\beta \epsilon^{1/2}), 
\\
\,\,\,\qquad\qquad\qquad\qquad\text{for}\, x\leq b(1-\beta \epsilon^{1/2}) 
\end{cases}
\eea


In the above, we have assumed $\epsilon \ll 1$ and anticipated that the 
distance scale to be$ \it{O}(\epsilon^{1/2})$, so that the variational 
parameters $\alpha$ and $\beta$ are numbers of $\it{O}(1)$. Note the 
matching occurs near the turning points in the classical regions for the 
corresponding Schr\"odinger equation. We need to evaluate 

\bea\label{18}
\lambda_{1}(\alpha , \beta )=\frac{\int^{\infty}_
{-\infty}e^{\psi(x)/2\epsilon}[-\epsilon \frac{\partial^{2}}
{\partial x^{2}}+\frac{U^{\prime}(x)^{2}}{4\epsilon}
+\frac{U^{\prime \prime}(x)}{2}]e^{\psi(x)/2\epsilon}dx}
{\int^{\infty}_{-\infty}e^{\psi(x)/\epsilon}dx}
\eea 

The calculation involves the following steps :

(i) Evaluating the normalization integral. We note that 
$\psi(x)\simeq U(0)- \frac{1}{2}U^{\prime \prime}(0)x^{2}$
for $x<1$ and the maximum value of $\psi(x)$ dominates the integral 
for $\epsilon \ll 1$. Thus, the normalization integral is
$e^{U_{0}/\epsilon}(2\pi \epsilon /U^{\prime \prime }(0))$.

(ii) The kinetic and potential energy term completely cancel 
$\psi(x)=U(x)$. Thus the integration in the numerator of Eq.(\ref{18}) 
involves integrating from $a(1-\alpha \epsilon^{1/2})$ to 
$\infty $ and from $b(1-\beta \epsilon^{1/2})$ to $-\infty $.

(iii) In the range of integration discussed in (ii), it is 
sufficient to approximate the potential $U(x)$ by a quadratic 
expression for $\epsilon \ll 1$. Terms like $U(a-\alpha a \epsilon^{1/2})$
and $U^{\prime }(a-a\alpha \epsilon^{1/2})$ need 
to be expanded to $\it{O}(\epsilon)$,
keeping in mind $U^{\prime }(a)=U^{\prime }(b)=0$. We thus obtain 
$\lambda_{1}(\alpha , \beta)$.

Minimizing $\lambda_{1}$ with respect to $\alpha $ and $\beta $ 
leads to the conditions: 

\bea\label{19}
a^{2}\alpha^{2}U^{\prime \prime }(a)=b^{2}\beta^{2}U^{\prime \prime }(b)=3
\eea
Thus, we find 

\bea\label{20}
\lambda_{1}=c[{\mid U^{\prime \prime}(0)\mid /U^{\prime \prime }(a)}^{1/2}
e^{-\triangle_{a}/\epsilon}+ {\mid U^{\prime \prime }(0)\mid /U^{\prime \prime }
(b)}^{1/2}e^{-\triangle_{b}/\epsilon}]
\eea 

with $\triangle_{a,b}=U(0)-U(a,b)$ and 

\bea\label{21}
c=\frac{1}{6}\frac{e^{3/2}}{(6\pi )^{1/2}}\simeq 0.17
\eea

For the WKB result of Caroli et al., $c= 1/2\pi \simeq 0.159 $.

\begin{figure}[t]
\begin{center}
\includegraphics[scale=0.4]{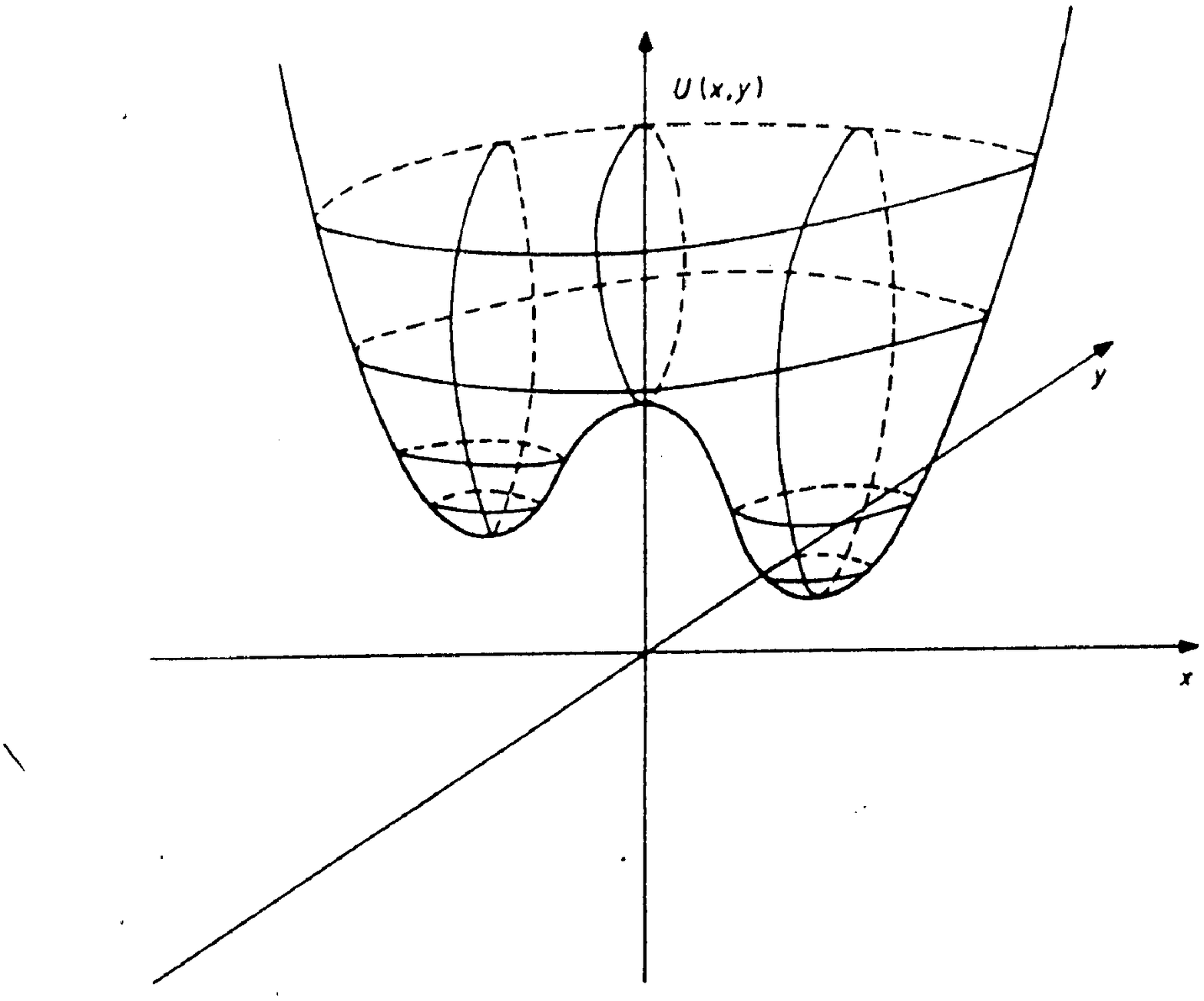}
\caption{\small{The general bistable potential in two dimensions.}}
\end{center}
\end{figure}
To end this section, we consider generalization of the above treatment
(Bhattacharjee and Banerjee, 1989)
to higher dimensions. In two dimensions, the bistable potential $U(x,y)$
has the form shown in fig.(2).
The important assumption about $U(x,y)$ is that there exists a most 
probable escape path (the instanton trajectory) connecting the two wells.
In this case, we can always choose the axes such that the desired path
is the x-axis. About this path, $U(x,y)$ can be expanded as

\bea\label{22}
U(x,y)=U(x)+\frac{1}{2}W(x)y^{2}
\eea

In the above $U(x)$ is the one dimensional potential that we have
already discussed. The Hamiltonian, corresponding to Eq.(\ref{5}) can now 
be written as

\bea\label{23}
H=-\epsilon \Bigg (\frac{\partial^{2}}{\partial x^{2}}+ \frac{\partial^{2}}
{\partial y^{2}}\Bigg )+V(x,y)
\eea
where

\bea\label{24}
V(x,y)=\frac{1}{4\epsilon }\Bigg [\Bigg (\frac{\partial U}
{\partial x}\Bigg )^{2}+
\Bigg (\frac{\partial U}{\partial y}\Bigg )^{2}\Bigg ]-\frac{1}{2}
\Bigg (\frac{\partial^{2}U}{\partial x^{2}}+
\frac{\partial^{2}U}{\partial y^{2}}\Bigg )
\eea

The ground state with zero eigenvalue is $\exp \Big(-U(x,y)/2\epsilon \Big )$.
As before, our interest is in the first excited state.

We begin by noting that the first excited state will involve changes, 
mainly in the $x$-direction, retaining the $y$-dependence of the wavefunction
as $\exp \Big (-W(x)y^{2}/2\epsilon \Big )$ to a good approximation. 
The "stiffness" in the $y$-direction suggests a variant of the 
Born-Oppenheimer approximation. We solve the $y$-dependent part of 
the above Hamiltonian, treating the $x$-dependence as a parameter.
The resulting eigenvalue will be a function of $x$ and can be treated as an 
effective potential for the one-dimentional problem in $x$. 
Introducing Eq.(\ref{22})
into Eq.(\ref{23}), the $y$-dependent Hamiltonian can be written as

\bea\label{25}
H_{y}=-\epsilon \frac{\partial^{2}}{\partial y^{2}}+\frac{1}{4\epsilon}
(W^{2}+U^{\prime}W^{\prime}-W^{\prime \prime}\epsilon)y^{2}
\eea

The lowest eigenvalue is $\frac{1}{2}(W^{2}+U^{\prime}W^{\prime}
-\epsilon W^{\prime \prime})^{1/2}$ and we use this as an effective 
potential for the one dimensional problem in $x$. The function $W(x)$ 
is going to be slowly varying and we can expect $W^{\prime}/W$ and 
$W^{\prime \prime}/W$ to be small. This allows us a binomial expansion
of the lowest eigenvalue and allows us to write the Hamiltonian as

\ber\label{26}
H &=& -\epsilon \frac{d^{2}}{dx^{2}}+\frac{(U^{\prime})^{2}}{4\epsilon}
-\frac{U^{\prime \prime}}{2}+\frac{U^{\prime}W^{\prime}}{4W}
-\frac{\epsilon W^{\prime \prime}}{4W} \nonumber \\
&\simeq & \epsilon \Bigg [-\frac {d}{dx}+\frac{U^{\prime}}{2\epsilon}
+ \frac{W^{\prime}}{4W}\Bigg ]\Bigg [\frac{d}{dx}+\frac{U^{\prime}}{2\epsilon}
+ \frac{W^{\prime}}{4W}\Bigg ] \nonumber \\
& & {}
\eer

where in the second step, we have deliberately not matched the 
higher-order term $(W^{\prime}/W)^{2}$. We require the two lowest 
eigenvalues of $H$ in Eq.(\ref{26}). The lowest eigenvalue is of course 
known to be exactly zero and so it is the approximate determination 
of $\lambda_{1}$ which is our concern and that in accordance with
our previous analysis, this is simply the ground 
state of the supersymmetric partner

\bea\label{27}
\overline{H}=\epsilon \Bigg [\frac{d}{dx}+\frac{U^{\prime}}{2\epsilon}
+\frac{W^{\prime}}{4W}\Bigg ]\Bigg [-\frac{d}{dx}+\frac{U^{\prime}}{2\epsilon}
+\frac{W^{\prime}}{4W}\Bigg ]
\eea

We now take over the variational calculation for the one-dimensional 
case dicussed above. The only difference to be noted is that the 
non-normalizable function which gives zero eigenvalue is now

\bea\label{28}
\psi(x)=\exp(U(x)/2\epsilon) \exp \Bigg (\frac{1}{4}\int^{x}_{0}
\frac{W^{\prime}(x)}{W(x)}dx \Bigg )
\eea

For $\epsilon \ll 1$, the extra term yields the factor 
$\exp(\frac{1}{2}ln \frac{W(a)}{W(0)})$ for the range of integration
$a(1-\alpha \epsilon^{1/2})$ to $\infty$ and a similar term 
for the range $b(1-\beta \epsilon^{1/2})$ to $-\infty$, when one 
calculates the expectation value. This leads to the answer

\bea\label{29}
\lambda_{1}= c \Bigg [\Bigg (\frac{\mid U^{\prime \prime }_{0}\mid}
{U^{\prime \prime} _{a}} \Bigg )^{1/2}(W_{a}/W_{0})^{1/2}
e^{-\triangle_{a}/\epsilon}
+\Bigg  (\frac{\mid U^{\prime \prime}_{0}\mid }{U^{\prime \prime}_{b}}
\Bigg )^{1/2}
(W_{b}/W_{0} )^{1/2}e^{-\triangle_{b}/\epsilon}\Bigg ]
\eea

where the constant $c$ is once again 
$\frac{e^{3/2}}{6}\frac{1}{(6\pi )^{1/2}}$.

To end this section we would like to point out a formal 
analogy (Schneider et al., 1985) between a quantum problem with potential
$V(x)$ and the Langevin process.
In Eq.(\ref{5}), if we write $t=i\tau$, we get the 
Schr\"{o}dinger equation ($\epsilon =\hbar/2m,\widetilde{V}=\hbar V$)

\bea\label{30}
i\hbar \frac{\partial \phi}{\partial \tau}=
\frac{-\hbar^{2}}{2m}\frac{\partial^{2}\phi}{\partial x^{2}}
+\widetilde{V}(x)\phi
\eea
and the stationary states in the $\tau$-variable are characterized
by the eigenvalues $\lambda_{n}$. In the path integral formulation
of quantum mechanics, this equation corresponds to the measure

\ber\label{31}
d\mu (x,t)&\sim & e^{i/\hbar \int d\tau 
\Big [(m/2)(\frac{\partial x}{\partial \tau})^{2}-\widetilde{V}(x)
\Big ]}{\mathcal D}[x(\tau)]\nonumber\\
&\sim &e^{-\int \frac{dt}{2m\epsilon}\Big [
\frac{M}{2}(\frac{dx}{dt})^{2}+V(x)\Big ]}
{\mathcal D}[x(t)]\nonumber\\
& & {}
\eer

Turning to the Fokker-Planck equation, note that 
if at $t=0$, $x=0$, then $P(x,t_{0})=\delta(x-x_{0})$
and this fixes the coffecients $a_{n}$ in Eq.(\ref{7}) as
$a_{n}=\exp(\lambda_{n}t_{0})\phi_{n}(x_{0})/\phi_{0}(x_{0})$ leading to

\bea\label{32}
P(x,t)=\phi_{0}(x)\sum_{n}\frac{\phi_{n}(x_{0})}
{\phi_{0}(x_{0})}e^{-\lambda_{n}(t-t_{0})}\phi_{n}(x)
\eea
with $x=x_{0}$ at $t=t_{0}$. The two-time correlation follows from

\ber\label{33}
<x(t)x(t_{0})>&=&\int dx\int dx_{0}P(x,t\mid x_{0},t_{0})xx_{0}
P_{eq.}(x_{0})\nonumber\\
&=&\int dxdx_{0}xx_{0}\phi_{0}(x)
\sum^{\infty}_{n=0}\frac{\phi_{n}(x_{0})}{\phi_{0}(x_{0})}
e^{-\lambda_{n}(t-t_{0})}\phi_{n}(x)\phi^{2}_{0}(x_{0})\nonumber\\
&=&\sum^{\infty}_{n=0}e^{-\lambda_{n}(t-t_{0})}
\int dx_{0}\phi_{0}(x_{0})x_{0}\phi_{n}(x_{0})
\int dx \phi_{0}(x)x\phi_{n}(x)\nonumber\\
&=&\sum_{n}\mid <0\mid x\mid n>\mid^{2}
e^{-\lambda_{n}(t-t_{0})}\nonumber\\
& & {}
\eer

For $t\gg t_{0}$, only the lowest eigenvalue will 
contribute and hence

\bea\label{34}
<x(t)x(0)>\longrightarrow \mid <0\mid x\mid 0>\mid^{2}+
\mid <0\mid x\mid 1>\mid^{2}e^{-\lambda_{1}t}
\eea

This particular correlation function, this explores the 
lowest eigenvalue of the quantum problem. Correlation 
of composite variables will explore other eigenvalues
$\lambda_{n}$. We thus arrive at the following result:
If the spectrum of a quantum mechanical problem with
potential $\widetilde{V}(x)$ is to be computed, then 
it should be possible to do that by studying a Langevin process
with potential $U(x)$, where $U(x)$ and $V(x)$ are related as
in Eq.(\ref{6}). Integration of the stochastic differential equation
then allows one to compute correlation function $<x(t)x(0)>$ as
\bea\label{35}
<x(t)x(0)>=\lim_{T\to\infty}\frac{1}{T}
\int x(t^{\prime})x(t^{\prime}+t)dt^{\prime}
\eea

The important issue is to find $U(x)$ if $V(x)$
is given. If the ground state energy of $H=T+V$,
where $T$ is the kinetic energy is $\lambda_{0}$
(this will not be zero in general), then $U$ has to be found 
from

\bea\label{36}
\frac{1}{4\epsilon}U^{\prime}(x)^{2}
-(1/2)U^{\prime \prime}=V(x)-\lambda_{0}
\eea

If $\phi_{0}$ is the ground state of $H$, then

\bea\label{37}
U=-2\epsilon \ln \phi_{0}
\eea

\section{Suzuki Regime}
In the bistable situations described in the previous section,
the intermediate time zone where the probability distribution
acquires bimodal character is also particularly important. We 
consider the evolution of an initial probability distribution
$P(x,t=0)$ in the bistable potential $V(x)=-\frac{x^{2}}{2}
+\frac{x^{4}}{4}$. At $t=0$, we assume that the probability 
is calculated about $x=0$ in a manner which is almost like a 
delta function (a very narrow Gaussian in reality). As time 
evolves, the distribution is going to spread out and then develop
two peaks. The spreading is the initial time regime, while the intermediate
time regime, where the probability distribution acquires the two peaks
is called the Suzuki regime because of the interesting scaling 
observed by Suzuki (1977) in this time-zone. If the probality distribution
explores only the small-$x$ part of the available space, then in the 
initial time regime, we can approximate $V(x)$ as 
$\widetilde{V}(x)=-x^{2}/2$. Carrying out the transformation 
$P(x,t)=\phi(x,t)\exp(-\widetilde{V}/2\epsilon)$,we find

\bea\label{38}
\frac{\partial \phi}{\partial t}=\epsilon 
\frac{\partial^{2}\phi}{\partial x^{2}}-
\Big (\frac{x^{2}}{4\epsilon}+\frac{1}{2}\Big )
\phi =H\phi
\eea

The eigenvalues of the operator $H$ are clearly $-(n+1)$
with the eigenfunction $\phi_{n}=N_{n}\exp(-x^{2}/\epsilon )
H_{n}(x/2\epsilon )$, where $H_{n}(y)$ is the $n^{th}$
Hermite polynomial and $N_{n}$ is the normalization constant.
We can immediately write

\bea\label{39}
P(x,t)=\sum C_{n}e^{-(n+1)t}N_{n}e^{-x^{2}/\epsilon }
H_{n}(x/2\epsilon )
\eea

The constants $C_{n}$ are fixed by the initial form of 
$P(x,t)$. If we take the idealized form of $P(x,t=0)=\delta (x)$,
then it follows that

\ber\label{40}
P(x,t)&=& \sum e^{-(n+1)t}N_{n}e^{-x^{2}/\epsilon }
H_{n}(0)H_{n}(x/2\epsilon )\nonumber \\
&=& \frac{1}{[2\pi \epsilon (e^{2t}-1)]^{1/2}}
e^{ -\frac{x^{2}}{2\epsilon (e^{2t}-1)}}\nonumber \\
& & {}
\eer

It is obvious that the probability distribution spreads out
in time since the width of the distribution is given by 
$2\epsilon (e^{2t}-1)\simeq 2\epsilon e^{2t}$. This is the 
initial time-regime, where the distribution sees the potential 
as an unstable inverted oscillator. The validity of the result 
for times such that $\epsilon e^{2t}\ll 1$ (ie. the spread is not too
big) or $t\ll t_{0}\sim ln(1/\epsilon )$.

We now consider $t\gg t_{0}$ but much less than the Kramer's time. 
In this limit the system 'slides down' the potential and does not
encounter the diffusion term Fokker-Planck equation. We can now 
write 

\ber\label{41}
\frac{\partial P}{\partial t}&=& \frac{\partial }{\partial x}
\Big [ P(-x+x^{3})\Big ]\nonumber \\
&=& (-x+x^{3})\frac{\partial P}{\partial x}+(3x^{2}-1)P\nonumber \\
& & {}
\eer

Writing $e^{t}=t^{\prime }$, we have

\bea\label{42}
t^{\prime }\frac{\partial P}{\partial t^{\prime}}=
(-x+x^{3})\frac{\partial P}{\partial x}+(3x^{2}-1)P
\eea

From the method of characterisation, we can write down the solution as

\ber\label{43}
P(x,t)&=& \frac{1}{x(1-x^{2})}G \Big (x^{2}/(t^{\prime })^{2}
(1-x^{2}) \Big )\nonumber \\
&=& \frac{1}{x(1-x^{2})}G \Big ( \frac{e^{-2t}x^{2}}{1-x^{2}}\Big )\nonumber \\
& & {}
\eer

where $G(z)$ is an arbitrary function. As always G has to be obtained
from the initial condition and in this case that is provided 
by Eq.(\ref{40}). The initial condition also entails $x\ll 1$ and comparing 
with Eq.(\ref{40}), we can obtain

\bea\label{44}
P(x,t)=\frac{1}{x(1-x^{2})}\frac{1}{[2\pi \epsilon (e^{2t}-1)]^{1/2}}
\exp \Big ( -x^{2}/2(1-x^{2})\epsilon (e^{2t}-1)\Big )
\eea

\begin{figure}[t]
\begin{center}
\includegraphics[scale=0.7]{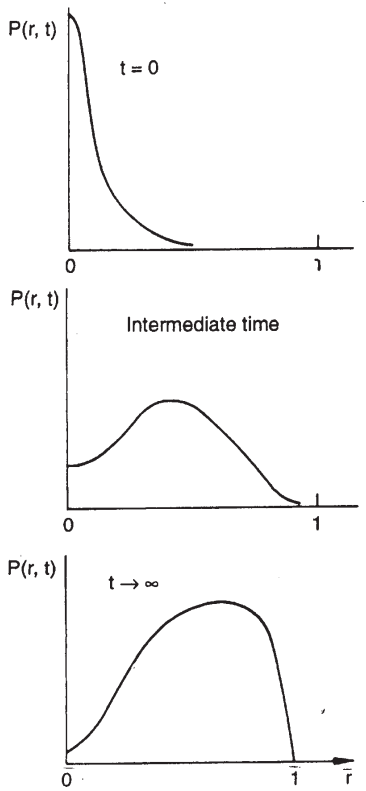}
\caption{\small{Time evolution of P(r,t)}}
\end{center}
\end{figure}

What emerges is the new time scale $\tau = \epsilon e^{2t}$, 
which determines when the probality distribution will acquire 
the two peaked structure. [The sequence of $P(r,t)$ for the
above function with $P(r,0)\simeq \delta (r)$ is shown in fig.(3).]
We will demonstrate that the above answer is exact if consider
a $N$-dimensional vector $x_{i}(t)$, $i=1,2,......,N $, 
with the stochastic time-evolution

\bea\label{45}
\frac{\partial x_{\alpha}}{\partial t} =
 -\gamma \frac{\partial V}{\partial x_{\alpha }}
+f_{\alpha }
\eea

with $V=-\frac{r^{2}}{2}+\frac{r^{4}}{4N}$ 
$r^{2}=\sum^{N}_{i=1}x_{i}^{2}$ and $f_{\alpha }$ is the random 
term with Gaussian correlation: 

\bea\label{46}
<f_{\alpha}(t_{2})f_{\beta }(t_{1})>=2\epsilon \delta (t_{1}-t_{2})
\eea

The fluctuation-dissipation relation holds with $\gamma = \epsilon $. 
The Fokker-Planck equation for $P({x_{\alpha }},t)$ is

\bea\label{47} 
\frac{\partial P}{\partial t}=\partial_{\alpha }
\Big (P\frac{\partial V}{\partial x_{\alpha }}\Big )
+\epsilon \frac{\partial^{2}P}{\partial x_{\alpha }\partial x_{\alpha }} 
\eea

For the spherically symmetric $V$, we expect $P=P(r,t)$ and 
thus in spherical polar coordinate system, Eq.(\ref{47}) reads

\bea\label{48}
\frac{\partial P}{\partial t}=\epsilon \Big (
\frac{\partial^{2}P}{\partial r^{2}}+
\frac{N-1}{r}\frac{\partial P}{\partial r}\Big )
+x_{\alpha }\Big (\frac{r^{2}}{N}-1\Big )
\frac{\partial P}{\partial x_{\alpha }}
+\Big [(1+2/N)r^{2}-N\Big ]P
\eea

We define $R=N^{-1/2}r$ and get

\bea\label{49}
\frac{\partial P}{\partial t}=\frac{\epsilon }{N}
\frac{\partial^{2}P}{\partial R^{2}}+
\frac{\epsilon (N-1)}{NR}
\frac{\partial P}{\partial R}+
R(R^{2}-1)\frac{\partial P}{\partial R}+
[(N+2)R^{2}-N]P
\eea

In the limit $N\gg 1$ and $\epsilon \ll 1$, we
drop the terms of $\it{O}(\epsilon )$,
$\it{O}(N^{-1})$ and $\it{O}(\epsilon /N)$, to
obtain

\bea\label{50}
\frac{\partial P}{\partial t}+
R(1-R^{2})\frac{\partial P}{\partial R}
=[(N+2)R^{2}-N]P
\eea

The method of characteristics now yields

\bea\label{51}
P(R,t)=\Big (\frac{1-R^{2}}{R^{2}}\Big )^{N/2}
\Big (\frac{1}{1-R^{2}}\Big )^{1+N/2}
G\Big (e^{2t}(1-R^{2})/R^{2}\Big )
\eea

Once again, this has to match smoothly to the small time
solution in order to determine the unknown function $G$.
Turning to Eq.(\ref{48}), the short-time solution is given by

\ber\label{52}
P(r\ll 1,t\ll t_{0})&=&
\Big [\frac{1}{2\pi \epsilon (e^{2t}-1)}\Big ]^{N/2}
e^{-\frac{r^{2}}{2\epsilon (e^{2t}-1)}}\nonumber \\
&=& \Big [\frac{1}{2\pi \epsilon (e^{2t}-1)}\Big ]^{N/2}
e^{-\frac{NR^{2}}{2\epsilon (e^{2t}-1)}}\nonumber \\
& & {}
\eer

The matching is done by the choice

\bea\label{53}
G\Big (e^{2t}(1-R^{2})/R^{2}\Big )=
\Bigg [\frac{R^{2}}{1-R^{2}}
\frac{N}{2\pi \epsilon (e^{2t}-1)}\Bigg ]^{N/2}
\exp \Bigg [-\frac{R^{2}}{1-R^{2}}
\frac{N}{2\epsilon (e^{2t}-1)}\Bigg ]
\eea

where $\exp(2t)$ has been replaced by $(\exp(2t)-1)$ 
since $\exp(2t)\gg 1$ and this replacement allows the
matching with Eq.(\ref{52}) to be implemented. This leads to the answer

\bea\label{54}
P(R,t)=\Bigg [\frac{1}{2\pi \epsilon (e^{2t}-1)}\Bigg ]^{N/2}
\Bigg (\frac{1}{1-R^{2}}\Bigg )^{1+N/2}
\exp \Bigg [-\frac{R^{2}}{1-R^{2}}
\frac{N}{2\epsilon (e^{2t}-1)}\Bigg ]
\eea

The discussion above is valid for a system of few
degrees of freedom and as such the principle area 
of application is laser physics, where the laser 
operates as a pump parameter crosses a threshold
value and then the exponential growth is checked 
by a cubic nonlinearity in the Langevin equation. 
The variable is the electric field which, for circularly 
polarised light, can be taken to be two-dimensional. 
The experimental measurements of the time-dependent
intensity clearly shows the existence of Suzuki regime.

Of greater interest is the exploration of Suzuki scaling
in the case of a field (Kawasaki et al., 1978; Bray, 1994),
ie., a function of space and time,
whose dynamics can be described by the Langevin equation

\bea\label{55}
\frac{\partial \phi (\vec{r},t)}{\partial t}
=-\frac{\delta F}{\delta \phi (\vec{r},t)}
+ f(\vec{r},t)
\eea

where $F$ is the "free energy" which can be written as
(in a $D$-dimensional space).

\bea\label{56}
F=\int \Bigg [-\frac{a}{2}\phi^{2}+
\frac{1}{2}({\bf{\nabla}} \phi )^{2}+
\frac{1}{2}(\phi^{2})^{2}\Bigg ]d^{D}x
\eea

In principle, $\phi $ may be a $N$-component vector, and
$\phi^{2}=\sum_{i=1}^{N}\phi^{2}_{i}$ and 
$({\bf{\nabla }}\phi )^{2}=\sum_{i,j}
(\partial_{i}\phi_{i})^{2}$.

The growth of order that we have been considering
crresponds to the parameter $a$ being positive. if 
we work with the Fourier components $\phi (\vec{k},t)$
of $\phi (\vec{r},t)$, then

\bea\label{57}
\frac{\partial \phi (\vec{k},t)}{\partial t}=
(a-k^{2})\phi (\vec{k},t)-\lambda \sum_
{\vec{k_{1}},\vec{k_{2}}}\phi (\vec{k_{1}},t)
\phi(\vec{k_{2}},t)\phi(\vec{k}-\vec{k_{1}}
-\vec{k_{2}},t)+f(\vec{k},t)
\eea 

The probability distribution $P({\phi(k)},t)$ satisfies

\ber\label{58}
\frac{\partial P(\phi(k),t)}{\partial t} &=&
\sum_{k^{\prime}}\frac{\partial }{\partial 
\phi(k^{\prime })}\Bigg [P(-a+(k^{\prime})^{2})
\phi(\vec{k^{\prime}},t)\nonumber\\
& &+ \lambda P\sum \phi(\vec{k_{1}},t)\phi(\vec{k_{2}},t)
\phi(\vec{k^{\prime}}-\vec{k_{1}}-\vec{k_{2}},t)
\Bigg ]\nonumber \\
& &+\sum_{k^{\prime}}\epsilon 
\frac{\partial^{2}P}{\partial \phi(k^{\prime})
\partial \phi(-k^{\prime })}\nonumber\\
& & {}
\eer

The growth occurs for those $k$-values which are smaller
than $a^{1/2}$, and our interest is in those wavenumbers alone. 
In the initial stages, where $\phi(x)$ is centred about 
$\phi =0$ and is small in magnitude, we can drop the cubic
term in the above equation and the time-development will occur 
according to $P(\phi(k),t)\sim \prod \limits_{k^{\prime }}
\exp \Bigg (-\mid \phi(k^{\prime})\mid^{2}/\epsilon e^{2rt}
\Bigg )$, where $r=a-(k^{\prime })^{2}$. In the intermediate
time-zone, it is the $\epsilon$-term in Eq.(\ref{58}) that needs to be 
dropped and we need to find the solution for $P(\phi(k),t)$ as
we did in Eq.(\ref{44}). However, now it is more complicated.

Simplification occurs if we look at the derivation of Eq.(\ref{44}) in
slightly different manner. Returning to that case and setting
$\epsilon =0$, implies in the Langevin picture solving the equation
$\dot {X}=X-X^{3}$. The solution is $X(t)=Ce^{t}/(1+C^{2}e^{2t})^{1/2}$ 
where C is a constant. If we call $X_{0}(t)=Ce^{t}=X(0)e^{t}$, then
clearly $X(t)=X_{0}(t)/[1+X^{2}_{0}(t)]^{1/2}$ and 
$X(0)=X(t)e^{-t}/[1-X^{2}(t)]^{1/2}$. We thus have a prescription 
for going to $X(0)$ from $X(t)$ and the evolution of $P(X,0)$ given 
as $e^{-X(0^{2})/\epsilon }$, will occur according to 
$P(X,t)\sim e^{-X^{2}/e^{2t}(1-X^{2})}$.

In the present case, this requires the solution of

\bea\label{59}
\frac{\partial \phi(k)}{\partial t}=(a-k^{2})\phi(\vec{k})
-\lambda \sum_{\vec{k_{1}},\vec{k_{2}}}\phi(\vec{k_{1}},t)
\phi(\vec{k_{2}},t)\phi(\vec{k}-\vec{k_{1}}-\vec{k_{2}},t)
\eea

The solution $\phi(k)$ proceeds according to standard perturbative
techniques. The simplification occurs in the large-time limit when
we can identify a leading term at every order. This allows the summation of
the perturbation series to yield

\bea\label{60}
\phi(\vec{r},t)=\phi_{0}(\vec{r},t)/
[1+\phi^{2}_{0}(\vec{r},t)]^{1/2}
\eea
with

\bea\label{61}
\phi_{0}(\vec{r},t)=\exp(t(a+\nabla^{2}))\phi(\vec{r})
\eea
The inverse transformation is

\bea\label{62}
\phi(\vec{r})=\exp(-t(a+\nabla^{2}))\phi(\vec{r})/
[1-\phi^{2}(\vec{r})]^{1/2}
\eea

In the absence of the diffusion term the probability
follows a Liouville equation
which is a conservation law for the probability. Hence, if the initial 
distribution is known, then the distribution at any time can be
obtained by the inverse transformation shown in Eq.(\ref{62}).

The evolution of this probability distribution can be pictured 
as follows. In the initial period, fluctuations everywhere 
start growing rapidly and at the same time diffuse over the distance
$(2rt)^{1/2}$, within which scale the fluctuations are strongly
correlatad. Saturation sets in as the value of $\phi^{2}(\vec{r})$
at any point approaches unity. After a while, in the language of 
magnetism, the system breaks up into domains of size 
$(2rt)^{1/2}$ with the saturation magnetisation at$+1$ or $-1$.

Finally, we would like to point out connection with a field theory
by writing the measure analogus to Eq.(\ref{31}) as 

\bea\label{63}
d\mu \sim \exp(-S[\phi ]) \prod_{\alpha =1}{\mathcal D}\phi_{\alpha}(x,t)
\eea
where
\ber\label{64}
S[\phi]&=&1/\hbar \int^{\infty}_{-\infty}{\mathcal H}dt \nonumber\\
&=& \int^{\infty}_{-\infty}dt\int d^{D}x\Big [\frac{
\dot{\phi_{\alpha}}\dot{\phi_{\alpha}}}{4\epsilon}
+V(\phi_{\alpha})\Big ]\nonumber\\
& & {}
\eer
with the potential $V(\phi_{\alpha})$ coming from

\bea\label{65}
V[\phi_{\alpha}]=\frac{1}{4\epsilon}
\frac{\delta U}{\delta \phi_{\alpha}}
\frac{\delta U}{\delta \phi_{\alpha}}-
\frac{1}{2}\frac{\delta^{2}U}
{\delta \phi_{\alpha}\delta \phi_{\alpha}}
\eea
With $t=i\tau$, the action is that for a quantum field theory
with the action $S_{q}$, where

\bea\label{66}
S_{q}=\int d\tau d^{D}x \Bigg [-\frac{1}{4\epsilon}
\frac{d\phi_{\alpha}}{d\tau}\frac{d\phi_{\alpha}}{d\tau}
+V(\phi_{\alpha})\Bigg ]
\eea

If $U=(1/2) m^{2}\phi_{\alpha}\phi_{\alpha}+
(1/2) ({\bf{\nabla \phi_{\alpha}}})^{2}$, ie., the Langevin potential
is quadretic, Eq.(\ref{65}), prescribes

\bea\label{67}
V[\phi_{\alpha}]=\frac{1}{4\epsilon}\Bigg [
m^{4}\phi_{\alpha}\phi_{\alpha}-\frac{m^{2}}{2}
-2m^{2}\phi_{\alpha}\frac{\partial}{\partial x_{\beta}}
\frac{\partial \phi_{\alpha}}{\partial x_{\beta}}
+\Bigg (\frac{\partial^{2}\phi }{\partial x_{\beta}\partial x_{\beta}}
\Bigg )^{2}\Bigg ]
\eea

Apart from constants,

\bea\label{68}
S_{q}=\int d\tau d^{D}x \Bigg [-\frac{1}{4\epsilon}
\frac{d\phi_{\alpha}}{d\tau}
\frac{d\phi_{\alpha}}{d\tau}+
\frac{m^{2}}{2\epsilon}\frac{\partial \phi_{\alpha}}
{\partial x_{\beta}}\frac{\partial \phi_{\alpha}}
{\partial x_{\beta}}+\frac{m^{4}}{4\epsilon}
\phi_{\alpha}\phi_{\alpha}+
\Bigg (\frac{\partial^{2}\phi}{\partial x_{\beta}\partial x_{\beta}}
\Bigg )^{2}\Bigg ]
\eea
which is a quadratic action and can be easily handled.

It is when $U$ is non trivial that one can generate non trivial
$S_{q}$ and what would be interesting is to consider a nontrivial
$S_{q}$(e.g., the one corresponding to the decay of the false vacuum
(Coleman,1977))
and see if the corresponding Langevin dynamics can shed light 
on the quantum problem.

\section{A Class of Time-Dependent Potentials}

In this section, we will deal with potentials which are
time-dependent but allow for the establishment of a final 
equilibrium state. This will be defferent from the time-dependences 
which are generally studied and have led
to a wealth of interesting phenomena. These include the cases
of diffusion over a barrier in the presence of harmonic force
and diffusion over a fluctuating barrier. The hallmark of the former
situation is the phenomenon of stochastic resonance
(Benzi et al., 1981; B\"{u}ttiker and Landauer, 1982; 
McNamara et al., 1988), where
the signal-to-noise ratio of the response to an applied force
displays a local maximum as a function of frequency. In the case of
fluctuating barriers
(Doering and Gadoua, 1992; Maddox, 1992; Z\"{u}rcher 
and Doering, 1993; Pechukas and H\"{a}nggi, 1994), the discovery that 
the mean first passage time
has a minimum as a function of the correlation time characterising 
the fluctuation has prompted a wide variety of investigations.
What we would like to present here is a toy model for yet another
kind of phenomenon-the global optimisation principle
(Doye et al., 1999; Hunjan and Ramaswamy, 2002) on an 
evolving potential energy landscape. In such problems one
is interested in finding the minima of a multidimensional 
potential energy surface which constitutes the energy landscape
in problems such as protein folding or the finding the 
ground state configuration of atomic or molecular clusters.
An interesting observation in this context is that 
of Hunjan et al.(to be published),
who have shown that a continuously and adiabatically varying 
potential assists approach to desired configuration at
$t \rightarrow \infty $ by avoiding trapping in local minima.

We want the final $(t= \infty )$ potential to be $V_{2}(x)$. 
We start from a different function and consider the time-dependent
potential

\bea\label{69}
V(x,t)=V_{2}(x)+b[V_{1}(x)-V_{2}(x)]\exp(-\lambda t)
\eea  
which has the form $b[V_{1}(x)-V_{2}(x)]$ at $t=0$
and evolves to $V_{2}(x)$ at $t=\infty$. For $b=1$, $V_{1}(x)$
is the initial shape which evolves to $V_{2}(x)$. Our 
goal is to study the approach to equilibrium in such a system.
It is clear that as $t \rightarrow \infty $ and $V \rightarrow V_{2}$,
there is an equilibrium probability distribution 
$P_{eq}=\exp(-V_{2}/\epsilon )$ at $t=\infty $ for 
the Fokker-Planck equation.

To study the onset of equilibrium when V is of the form
shown in Eq.(\ref{69}), we make the usual substitution

\ber\label{70}
P(x,t)&=&\phi(x,t)e^{-V(x,t)/2\epsilon}\nonumber\\
&=&\phi(x,t)e^{-V_{2}(x)/2\epsilon}
e^{-b\triangle V e^{-\lambda t}/2\epsilon}\nonumber\\
& & {}
\eer
where 

\bea\label{71}
\triangle V=V_{1}(x)-V_{2}(x)
\eea
leading to

\bea\label{72}
\frac{\partial \phi}{\partial t}
=H_{0}\phi +H_{1}(t)\phi
\eea 
with

\bea\label{73}
H_{0}\phi =\epsilon \frac{\partial^{2}\phi}
{\partial x^{2}}+\frac{1}{2}V^{\prime \prime}_{2}\phi
-\frac{1}{4\epsilon }(V^{\prime}_{2})^{2}\phi
\eea
and

\bea\label{74}
H_{1}=b\Bigg [\frac{\triangle V^{\prime \prime}}{2}
-\frac{\lambda}{2\epsilon }\triangle V-
\frac{V^{\prime}_{2}(\triangle V)^{\prime}}{2\epsilon}
\Bigg ]e^{-\lambda t}-b^{2}\frac{(\triangle V^{\prime})^{2}}
{4\epsilon}e^{-2\lambda t}
\eea

In the above equation prime denotes differentiation with
respect to $x$. The eigenvalues of $H_{0}$ are non-positive
and we can write

\bea\label{75}
H_{0}\psi_{n}=-E_{n}\psi_{n},\ with \ E_{n}\ge 0
\eea 

The ground state $\psi_{0}$ has a space-independent part
$V_{0}(t)$. We will separate out this part and write 
the general solution of Eq.(\ref{72}) as

\bea\label{76}
\phi(x,t)=\sum c_{n}(t)e^{-E_{n}t+\int^{t}V_{0}(t^{\prime})dt^{\prime}}
\psi_{n}
\eea

The usual techniques of Dirac's time dependent perturbation theory
lead to

\bea\label{77}
\dot{c_{n}}(t)=\sum_{m}c_{m}(t)<m\mid H_{2}\mid n>\exp[-(E_{m}-E_{n})t]
\eea
where $H_{2}=H_{1}-V_{0}$. Perturbation theory proceeds by expanding

\bea\label{78}
c_{n}(t)=c_{n0}(t)+bc_{n1}(t)+b^{2}c_{n1}(t)+............
\eea

Introducing it in Eq.(\ref{77}) yields

\bea\label{79}
\dot{c_{n0}}=0
\eea

\bea\label{80}
\dot{c_{n1}}(t)=\sum_{m}c_{m0}(t)<m\mid H\mid n>\exp[-(E_{m}-E_{n})t]
\eea

\bea\label{81}
\dot{c_{n2}}(t)=\sum_{m}c_{m1}(t)<m\mid H\mid n>\exp[-(E_{m}-E_{n})t]
\eea
and so on.
We see immediately that $c_{n0}=constant$, independent of time.
Consequently, they are determined by the state of the system
at $t=0$.

We focus on the where the final and initial potentials
have qualitatively different structures. Our $V_{2}$ will 
be a double well potential, while our $V_{1}$ is the usual 
single well potential. Approach to equilibrium in the double 
well potential is strongly delayed, as we have seen in sec. 2, 
by the Kramers' time which is a long time scale coming from 
the possibility of noise induced hopping from one minimum 
to another. Thus $V_{2}(x)=-\frac{x^{2}}{2}+\frac{x^{4}}{4}$,
while $V_{1}(x)=x^{2}$.With this choice

\bea\label{82}
H_{0}\phi=-\epsilon \phi^{\prime \prime}+
\Bigg (\frac{(V^{\prime}_{2})^{2}}{4\epsilon}
-\frac{V^{\prime \prime}_{2}}{2}\Bigg )\phi
\eea

\ber\label{83}
H^{\prime}&=&\frac{3}{2}(1-x^{2})-(1/4\epsilon)x^{2}
(3-x^{2})(1+x^{2})\nonumber\\
&-&(\lambda /8\epsilon)x^{2}(6-x^{2})
+(1/4\epsilon)x^{2}(3-x^{2})^{2}(1-e^{\lambda t})\nonumber\\
& & {}
\eer

The eigenvalues spectrum of $H_{0}$ is characterized, as we have seen before,
by a set of close doublets as its low lying states. The separation 
within the doublet is exponentially small, while that between two doublets
is of $\it{O}(1)$. The ground state is $E_{0}=0$ and the first exited 
state is the ground state of the supersymmetric partner of
$\frac{(V^{\prime}_{2})^{2}}{4\epsilon}-\frac{V^{\prime \prime}_{2}}{2}$
and is exponentially small. The second exited state has an eigenvalue 
close to 2 and hence we can safely approximate the dynamics of the low 
lying states as that of a 2-atate system. If $\phi_{0}$ and $\phi_{1}$ 
are the two  states and $0$ and $\delta$ the eigenvalues, then 
we can write

\bea\label{84}
\phi(x,t)=c_{0}(t)\phi_{0}(x)+c_{1}(t)\phi_{1}(x)\exp(-\delta t)
\eea   
Note that $\delta^{-1}$ is the Kramers' time in the problem.
The dynamics of $c_{0}$ and $c_{1}$ is governed by

\bea\label{85}
\dot{c_{0}}=<\phi_{0}\mid H^{\prime}\mid \phi_{0}>e^{-\lambda t}
c_{0}(t)+<\phi_{0}\mid H^{\prime}\mid \phi_{1}>c_{1}(t)
e^{-\delta t}e^{-\lambda t}
\eea

\bea\label{86}
\dot{c_{1}}=c_{0}<\phi_{1}\mid H^{\prime}\mid \phi_{0}>
e^{\delta t}e^{-\lambda t}+
c_{1}<\phi_{1}\mid H^{\prime}\mid \phi_{1}>e^{-\lambda t}
\eea

Now $H^{\prime}$ is even and hence 
$<\phi_{1}\mid H^{\prime}\mid \phi_{0}>=0$, which 
decouples $c_{0}$ and $c_{1}$ and we can easily integrate 
the above equations. 
If we drop terms like $\exp(-2\lambda t)$, we get

\bea\label{87}
\dot{c_{1}}=c_{1}<\phi_{1}\mid \widetilde{H}^{\prime}\mid \phi_{1}>
\exp(-\lambda t)
\eea 
where

\bea\label{88}
\widetilde{H}^{\prime}=(3/2)(1-x^{2})+(2x^{2}/4\epsilon)
(3-x^{2})(1-x^{2})-(\lambda/8\epsilon)x^{2}(6-x^{2})
\eea

The dominant contributions to both 
$<\phi_{1}\mid \widetilde{H}^{\prime}\mid \phi_{1}>$
and $<\phi_{0}\mid \widetilde{H}^{\prime}\mid \phi_{0}>$ comes
from near $x\simeq 1$.
The small difference between the two matrix elements 
comes from the term $(3/2)(1-x^{2})$ in $H^{\prime}$,
which is maximum near $x=0$ and at that point 
$\phi_{0}\not\simeq 0$ but $\phi_{1}\simeq 0$.
After a set of straightforward manipulations, we see that

\bea\label{89}
P(x,t)=P_{eq}(x) [1+f(x)\exp(-\delta_{eff}t)]
\eea
where

\bea\label{90}
\delta_{eff}=\delta +\frac{3(1-e^{\lambda t})}{2\lambda t}\alpha
\eea
where $\alpha$ is a small number of the prder of $\delta$.

The approach to equilibrium is now through a modified
Kramers' time which is obtained from $\delta^{-1}_{eff}$.
Clearly $\delta_{eff}>\delta$ and hence the new equilibration
time is going to be smaller. Thus, in the toy model, we see
a faster approach to equilibrium , which was the desired goal.

\end{document}